\documentclass[aps,twocolumn,superscriptaddress,english,10pt,nofootinbib]{revtex4}

\usepackage{color}
\usepackage{graphicx}
\usepackage{latexsym}
\usepackage{dcolumn}
\usepackage{bm}

\usepackage{hyperref}
\usepackage{amsmath}
\usepackage{parskip}
\usepackage{adjustbox}
\usepackage{mathtools}
\usepackage{empheq}
\usepackage[normalem]{ulem}
\hypersetup{
    colorlinks=true,
    linkcolor=blue,
    filecolor=magenta,      
    urlcolor=blue,
    pdftitle={KRandGravPV},
    pdfpagemode=FullScreen,
    }

\def\@Aboxed#1&#2\ENDDNE{%
  \settowidth\@tempdima{$\displaystyle#1{}$}%
  \addtolength\@tempdima{\fboxsep}%
  \addtolength\@tempdima{\fboxrule}%
  \global\@tempdima=\@tempdima
  \kern\@tempdima
  &
  \kern-\@tempdima
  \fcolorbox{red}{yellow}{$\displaystyle #1#2$}
}

\newlength\dlf

\def\tV{\tilde{\mathcal{V}}}

\def\Ve{\mathcal{V}_e}
\def\Vm{\mathcal{V}_m}
\def\tB{\tilde{B}}
\def\tVe{\tilde{\mathcal{V}}_e}
\def\tVm{\tilde{\mathcal{V}}_m}
\usepackage[title]{appendix}

\begin{document}
\title{The Kalb-Ramond field and Gravitational Parity Violation}
\medskip\
\author{Tucker Manton} 
\email[Email: ]{ tucker\_manton$@$brown.edu}
\affiliation{Department of Physics,
Brown University, Providence, RI 02912, USA}
\author{Stephon Alexander}%
\email[Email: ]{ stephon\_alexander$@$brown.edu}
\affiliation{Department of Physics,
Brown University, Providence, RI 02912, USA}

\date{\today}% It is always \today, today,

\begin{abstract}
Gravitational wave physics can probe theories that extend beyond General Relativity. Motivated by the attention on the Kalb-Ramond field as a dark matter candidate, in this work, we study a parity violating dimension four operator which couples the dual Riemann curvature to the 2-form field. After mapping the equations of motion for the right- and left-handed gravitational wave amplitudes to the novel parameterization presented in \cite{Jenks:2023pmk}, we discuss constraints on the model parameters in light of the coincident electromagnetic/gravitational wave signal of GW170817. 
\end{abstract}

\pacs{Valid PACS appear here}% PACS, the Physics and Astronomy
\maketitle
%\onecolumngrid
%=================================================================================================================================

\section{Introduction}\label{sec:Intro}
The advent of gravitational wave (GW) astronomy has ushered in a new era of modern physics, pioneered by the growing catalog of observations from the LIGO-Virgo-Kagra collaboration \cite{LIGOScientific:2014pky,VIRGO:2014yos,LIGOScientific:2020ibl,KAGRA:2021vkt} with recent data supporting a stochastic GW background from the pulsar timing array collaborations \cite{NANOGrav:2023gor,NANOGrav:2023hvm,EPTA:2023fyk,Xu:2023wog,Reardon:2023gzh}. GW signals provide a rich compliment to electromagnetic signals of astrophysical phenomena \cite{LIGOScientific:2017zic}, as well as offering a new arena for testing General Relativity (GR) in strong field regimes \cite{LIGOScientific:2021sio}. An immense outstanding question in the field is not if GR will fail to describe a phenomenon, but when and how. In particular, observing gravitational parity violation would be a smoking gun for physics beyond GR, as the Einstein-Hilbert term alone produces no parity violation. Parity violating gravitational theories have gathered interest in recent decades for a variety of reasons. Observing parity violation in CMB polarization measurements was considered as far back as 1998 \cite{Lue:1998mq}.

Dynamical Chern-Simons theory (dCS) is likely the most studied parity violating model for gravity. In addition to the two usual massless spin-2 degrees of freedom in GR, dCS includes a pseudoscalar field $\phi$ with a canonical kinetic term coupled to the Pontryagin density $\tilde{R}_{\mu\nu\rho\sigma}R^{\mu\nu\rho\sigma}$, where $\tilde{R}_{\mu\nu\rho\sigma}=\frac{1}{2}\varepsilon_{\mu\nu\alpha\beta}R^{\alpha\beta}_{ \ \ \ \rho\sigma}$ is the dual Riemann tensor, which arises in many contexts \cite{Yunes:2009hc} (see \cite{Alexander:2009tp} and references therein for a review). Interesting physics follows if the dCS pseudoscalar acts as the inflaton. This includes a preheating mechanism \cite{Alexander:2014bsa}, leptogenesis \cite{Alexander:2004us}, as well as the potential for parity violation to be detected in large-scale structure via galaxy correlations \cite{Creque-Sarbinowski:2023wmb}. Perturbations to rotating black holes were also studied in dCS theory recently in \cite{Wagle:2023fwl}, and in \cite{Nojiri:2019nar}, the propagation of gravitational waves in dCS were compared to standard $f(R)$ modified gravity theories. The dCS coupling was also studied in the context of the string axiverse in \cite{Yoshida:2017cjl}, where the authors considered a late-time scenario where axions are the dominant dark matter component, and discussed constraints on the coupling.

There are numerous other parity violating theories on the market. One such class is the ghost-free higher derivative models \cite{Crisostomi:2017ugk,Qiao:2021fwi}, which are characterized by Lagrangians of the form $\mathcal{L}(g,\partial g,\partial^2g)$, and individual terms are contractions of the metric, Riemann, and the Levi-Civita tensors. Higher derivative theories are generically plagued by Ostrogradsky instabilities \cite{Woodard:2015zca}, therefore such theories can at best be valid up to the mass scale of the ghost degrees of freedom from an effective field theory standpoint. Adjacent to these are the ghost-free scalar tensor models, such as those discussed in \cite{Nishizawa:2018srh}. Here, derivatives of a scalar field couple to contractions of the Riemann and Levi-Civita tensors. As in the higher order derivative theories, constraints on the Lagrangian coefficients must be imposed in order to avoid Ostrogradsky ghosts. There is also a parity violating addition to the Ho\^rava-Lifshitz approach to quantum gravity \cite{Horava:2009uw,Wang:2017brl}, which was studied in the context of gravitational wave physics first in \cite{Takahashi:2009wc} and later in \cite{Zhu:2013fja}. A different construction was proposed in \cite{Takahashi:2022mew}; here, the authors employed conformal transformations of the metric where the conformal factor is a function of the Pontryagin density. This produces a parity violating effect in the Einstein frame, which was recently used to study modifications of the Kerr metric \cite{Tahara:2023pyg}.

Alternatively, one can relax the standard assumptions of metricity (the so-called teleparallel gravity theories)  and vanishing torsion \cite{Conroy:2019ibo, Iosifidis:2020dck, Freidel:2005sn,Mavromatos:2023wkk,Shapiro:2014kma,Szczachor:2012hh}. In theories of this type, parity violating couplings between the Levi-Civita and either non-metricity or torsion tensors are expected on general grounds, which has been known since the late 1970's \cite{Hojman:1980kv}. The authors of \cite{Jenks:2023pmk} analyzed many of these examples, notably dynamical Chern-Simons, ghost-free scalar tensor gravity, (symmetric) teleparallel, and Ho\^rava-Lifshitz.

A common feature of many of these models is that the parity violating operators have dimension greater than four. Returning back to dCS, the interaction is a dimension five operator, and  therefore necessitates a mass scale,
\begin{equation}
    \mathcal{L}_{\text{dCS}}=\frac{\phi}{m_{\text{CS}}}R\tilde{R},
\end{equation}
implying that its potential affect on gravitational parity violation is suppressed. Moreover, dCS being quadratic in curvature has the consequence that the field equations are higher than second order \cite{PhysRevD.86.124031}. It is therefore well worth exploring the possibility of lower dimensional operators which will not come with an (inverse) mass scale and, idealistically, be manifestly safe from instabilities arising from higher derivatives. Studying such an interaction is precisely the purpose of this note. 

The presence of the Levi-Civita symbol contracted with the Riemann tensor results in an overall sign difference between the equations of motion for gravitational wave amplitudes when written in the circular basis. That is, for a gravitational wave propagating in the $z$-direction with polarizations in the $xy$-plane, we write 
\begin{equation}
    h_{\mu\nu}=\frac{1}{\sqrt{2}}\begin{pmatrix}
        0&0&0&0\\
        0&-(h_{\text{L}}+h_{\text{R}})&i(h_{\text{L}}-h_{\text{R}})&0 \\
        0&i(h_{\text{L}}-h_{\text{R}})&h_{\text{L}}+h_{\text{R}}&0 \\
        0&0&0&0 
    \end{pmatrix},
\end{equation}
where the right- and left-handed strains are related to the usual $+,\times$ amplitudes by $h_{\text{R,L}}=\frac{1}{\sqrt{2}}(h_+\pm ih_\times).$ Varying the Einstein-Hilbert Lagrangian with respect to each amplitude results in simple wave equations for each. Parity violation will then follow provided an additional interaction term sources the left- and right-handed amplitudes with a different sign, say
\begin{equation}
    \begin{split}
        \Box h_{\text{L}}&=+\Xi,\\
        \Box h_{\text{R}}&=-\Xi,
    \end{split}
\end{equation}
for some $\Xi$ that is  an expression involving additional matter fields and perhaps their derivatives\footnote{The authors of \cite{Contaldi:2008yz} approached the problem by considering the right- and left-handed waves as propagating with different effective Newton constants. This affects the vacuum fluctuations during inflation and the parity violation would translate into anomalous CMB polarization. }.  As usual, we define the graviton as
\begin{equation}
    g_{\mu\nu}=\bar{g}_{\mu\nu}+h_{\mu\nu},
\end{equation}
where $\bar{g}_{\mu\nu}$ in this work will be the flat FLRW metric in conformal coordinates,
\begin{equation}
    d\bar{s}^2=a(\eta)^2\big(-d\eta^2+dx^2+dy^2+dz^2\big).
\end{equation}

Parity violating effects in the gravitational sector generically demand a non-minimal coupling to matter, such as the pseudoscalar in dCS theory. It is interesting to entertain the possibility that such matter belongs to a dark sector. Very recently, the Kalb-Ramond (KR) two-form field $B_{\mu\nu}$ was studied as a dark matter candidate \cite{Capanelli:2023uwv}. The authors showed that the KR field and Kalb-Ramond-like-particles (KRLPs) can be produced via freeze-in and freeze-out mechanisms and can account for the relic density of dark matter today. Others have considered the KR field as a portal between the SM and dark sectors \cite{Dashko:2018dsw,Dick:2020eop}. It has also been studied in the context of gravitational lensing \cite{Chakraborty:2016lxo}, stellar physics \cite{Chakraborty:2017uku}, and more generally in early universe scenarios \cite{BeltranAlmeida:2018nin,Almeida:2019xzt} including leptogenesis \cite{deCesare:2014dga}, as well as late time cosmic acceleration \cite{Orjuela-Quintana:2022jrg,BeltranAlmeida:2019fou}.  

The KR 2-form is one of three fields in the spectrum of bosonic string theories \cite{PhysRevD.9.2273,Ogievetsky:1966eiu}, also arising in a $(1,0)\oplus (0,1)$ representation of the Lorentz group \cite{Ivanov:2016lha}. The KR field is additionally relevant in the context of axion physics; in the non-interacting (massless) case, the KR field strength can be dualized to obtain a pseudoscalar often referred to as the KR axion \cite{Svrcek:2006yi} (see also \cite{Smailagic:2001ch} for an earlier discussion).

Inspired in particular by the recent attention to the KR field as a dark matter candidate in models with broken gauge invariance \cite{Capanelli:2023uwv}, we consider parity violating effects due to a coupling between the dual-Riemann tensor and the KR;
\begin{equation}\label{Lint}
    \mathcal{L}_{\text{int}}=-\frac{\xi}{2}\tilde{R}_{\mu\nu\rho\sigma}B^{\mu\nu}B^{\rho\sigma}.
\end{equation}
The KR field carries mass dimension one, therefore in difference from the dCS coupling, the constant $\xi$ is dimensionless. This term was written down in \cite{Altschul:2009ae} in the context of Lorentz-violating vector and scalar theories, which we will discuss further in the conclusion. However, to the best of our knowledge, (\ref{Lint}) has not been studied in the context of parity violation\footnote{In \cite{Maity:2004he}, the authors considered couplings between the KR and Maxwell fields in the context of parity violation, while in \cite{Majumdar:2001ff,Majumdar:1999jd}, a similar construction was explored using the KR axion. }.

This paper is organized as follows. In Sec. \ref{sec:KRfeatures}, we discuss aspects of the free KR Lagrangian, including a symmetry breaking process that leads to a simplified scenario where the field components are constant. In Sec. \ref{sec:IntFeatures}, we discuss properties of the interaction (\ref{Lint}) in more detail before introducing the full theory in Sec. \ref{sec:FullTheory}.  We then focus on the perturbations in Sec. \ref{sec:Parameterization}, where we make contact with the novel, parity violating gravitational wave parameterization recently presented in \cite{Jenks:2023pmk}. Here we find the solutions to our model in an arbitrary FLRW cosmology. In Sec. \ref{sec:Constraints}, we discuss constraints on the model parameters from GW170817 and the GWTC-3 data set, before concluding in Sec. \ref{sec:Discussion}.
%=================================================================================================================================
\section{Features of the KR Lagrangian}\label{sec:KRfeatures}
%=================================================================================================================================

The KR field is described by the Lagrangian \cite{PhysRevD.9.2273} 
\begin{equation}\label{LB}
    \mathcal{L}_{\text{B}}=\frac{1}{12}H_{\mu\nu\rho}H^{\mu\nu\rho}-V(B),
\end{equation}
where 
\begin{equation} 
H_{\mu\nu\rho}=\partial_{[\mu}B_{\nu\rho]}=\partial_\mu B_{\nu\rho}+\partial_\nu B_{\rho\mu}+\partial_\rho B_{\mu\nu}
\end{equation}
is the KR field strength and $V(B)$ is some potential. When $V=0$, the theory enjoys a gauge symmetry with transformation\footnote{We use the anti-symmetrization convention $A_{[\mu}B_{\nu]}=A_\mu B_\nu-A_\nu B_\mu$.}  
\begin{equation}\label{gauge1}
    B_{\mu\nu}\rightarrow B_{\mu\nu}+\partial_{[\mu}A_{\nu]} 
\end{equation}
for an arbitrary vector $A_\mu(x)$, corresponding to a shift in $B$ by an exact 2-form. There is an additional transformation involving a scalar field, 
\begin{equation}\label{gauge2}
    A_\mu\rightarrow A_\mu+\partial_\mu\lambda.
\end{equation}
Thus although $A_\mu$ has four components, only three are independent. The subsidiary gauge transformation (\ref{gauge2}) is a shift involving an exact 1-form. As briefly mentioned in Sec. \ref{sec:Intro}, the gauge invariant theory is dual to the KR axion $\varphi$ via $\partial_\mu\varphi\sim\varepsilon_{\mu\nu\rho\sigma}H^{\nu\rho\sigma}$ \cite{Svrcek:2006yi,Smailagic:2001ch} and as is usual in axion physics, couplings to matter fields are generically parity violating. The situation is different here, however, due to presence of the potential and the non-minimal coupling to the dual Riemann tensor.

The gauge invariance (\ref{gauge1}), (\ref{gauge2}) being broken explicitly by the potential $V$ and the interaction (\ref{Lint}) has analogies to the Proca theory of a massive vector boson, which is known to be dual to the massive KR field in certain regimes \cite{Hell:2021wzm,Capanelli:2023uwv}. Regarding interactions with the gravitational waves, we will see that a simple and interesting scenario occurs when the KR field's magnitude settles at a non-zero, constant (classical) value, $B^2=B_{(0)}^2$. This can be modeled by a potential of the form 
\begin{equation}\label{potential}
    V(B_{\mu\nu}B^{\mu\nu}-v^2).
\end{equation}
When the KR field relaxes to the minimum, we have
\begin{equation}
    v^2\rightarrow \bar{g}^{\mu\alpha}\bar{g}^{\nu\beta}B_{\mu\nu}^{(0)}B_{\alpha\beta}^{(0)},
\end{equation}
where $\bar{g}^{\mu\nu}$ is the unperturbed metric. The explicit form of the potential (\ref{potential}) is not relevant for this work, beyond requiring that it possesses a nontrivial stable minimum\footnote{A potential of this form can arise rather naturally. In \cite{Assuncao:2019azw}, it was shown that an effective potential for the KR field possessing a minimum emerges from quantum corrections following couplings to fermions. }. In theories of this type, such potentials have the consequence that the physics can display (spontaneous) Lorentz violation, despite the Lagrange density being Lorentz invariant \cite{Altschul:2009ae}.

The KR field can be parameterized in a particularly simple form \cite{Altschul:2009ae,Maluf:2021eyu}. In general, a single KR field in a constant background value breaks the isotropy of the cosmology \cite{Maluf:2021eyu}, analogous to the case of a single vector field \cite{Golovnev:2008cf}. Global isotropy can be recovered if multiple vector fields take on constant values of equal magnitude orthogonal to each other, and we expect a similar story to hold here. We will save a treatment of such a scenario for future work, and will solely consider the effect of the interaction on gravitational waves approximating an isotropic cosmology. We therefore parameterize the background field as\footnote{Note in difference from \cite{Altschul:2009ae,Maluf:2021eyu}, we include the factor of $\tfrac{1}{\sqrt{2}}$ for convenience.} 
\begin{equation}\label{vev}
    B^{(0)}_{\mu\nu}=\frac{1}{\sqrt{2}}\begin{pmatrix}
        0&-\Ve a^2&0&0\\
        \Ve a^2 &0&0&0 \\
        0&0&0&\Vm a^2 \\
        0&0&-\Vm a^2 & 0
    \end{pmatrix},
\end{equation}
where $a(\eta)$ is the scale factor and $\Ve,\Vm$ are taken to be real numbers, and can be thought of as the ``electric'' and ``magnetic'' parts of the KR field, respectively. It then follows that the minimum of the potential is at 
\begin{equation}\label{minOfV}
    v=\sqrt{\Vm^2-\Ve^2},
\end{equation}
independent of the background evolution of the spacetime. It is also natural to expect a mass term for the KR field in this scenario. For the field to be frozen at the bottom of the minimum surrounding $v$, we require the mass to be slightly smaller than but on the order of $v$. This will guarantee that the symmetry breaking minimum exists while simultaneously ensuring the field is heavy enough such that no radial excitations occur, allowing us to treat the field as constant. However, as we look to make contact with observations in the late universe with this construction, this is not a requirement in early universe physics.

An alternative justification for considering a frozen KR background is by simply examining the background evolution in a flat FLRW spacetime without the interaction term. The field equation in this case is
\begin{equation}
 B''_{\mu\nu}+\delta_{\eta[\mu }B''_{\nu] \eta}+2\mathcal{H}\delta_{\eta[\mu}B'_{\nu]\eta}+a^2\frac{\partial V}{\partial B^{\mu\nu}}=0,
\end{equation}
where we've assumed $B_{\mu\nu}=B_{\mu\nu}(\eta)$ and $\mathcal{H}=a'/a$ is the Hubble parameter. As is standard in an expanding cosmological background, the presence of Hubble friction ultimately dampens the evolution of the unperturbed field until it rests at a constant value. This, of course, will be at the minimum of the potential where $\frac{\partial V}{\partial B^{\mu\nu}}=0.$

%=================================================================================================================================
\section{Features of the interaction}\label{sec:IntFeatures}
%=================================================================================================================================
The interaction (\ref{Lint}) is a specific case of a more general coupling 
\begin{equation}\label{Pterm}
    \mathcal{L}\sim R_{\mu\nu\rho\sigma}P^{\mu\nu\rho\sigma},
\end{equation}
since here, we can move the Levi-Civita symbol over to one of the $B$-fields and equivalently write  
\begin{equation}\label{Lint2}
    \mathcal{L}_{\text{int}}=-\frac{\xi}{2} R_{\mu\nu\rho\sigma}\tilde{B}^{\mu\nu}B^{\rho\sigma},
\end{equation}
where $\tilde{B}^{\mu\nu}=\frac{1}{2}\varepsilon^{\mu\nu}_{ \ \ \alpha\beta}B^{\alpha\beta}$. From (\ref{Pterm}), we see that in considering equations of motion, we have the derivative
\begin{equation}\label{RBtildeB}
    \frac{\partial\mathcal{L}}{\partial R_{\mu\nu\rho\sigma}}=P^{\mu\nu\rho\sigma},
\end{equation}
which implies that $P$ must inherit all of the symmetries of the Riemann tensor. That fact is not manifest as it is written in (\ref{Lint2}), therefore care must be taken when we derive the equations of motion covariantly. In particular, we have that 
\begin{equation}\label{FullP}
\begin{split} 
    P^{\mu\nu\rho\sigma}&=\frac{1}{3}\Big(\tB^{\mu\nu}B^{\rho\sigma}+B^{\mu\nu}\tB^{\rho\sigma}\Big) \\
    &+\frac{1}{6}\Big(\tB^{\mu\rho}B^{\nu\sigma}+B^{\mu\rho}\tB^{\nu\sigma}+\tB^{\rho\nu}B^{\mu\sigma}+B^{\rho\nu}\tB^{\mu\sigma}\Big).
    \end{split}
\end{equation}
It is tedious but straightforward to show that this indeed possesses all symmetries of the Riemann tensor,
\begin{equation}\label{syms}
    P_{\mu\nu\rho\sigma}=-P_{\nu\mu\rho\sigma}=-P_{\mu\nu\sigma\rho}=P_{\rho\sigma\mu\nu},
\end{equation}
along with the cyclic property
\begin{equation}
    P_{\mu\nu\rho\sigma}+P_{\mu\rho\sigma\nu}+P_{\mu\sigma\nu\rho}=0.
\end{equation} 
However, when contracted with the Riemann tensor, all of the terms combine such that 
\begin{equation}
    R_{\mu\nu\rho\sigma}P^{\mu\nu\rho\sigma}=R_{\mu\nu\rho\sigma}\tilde{B}^{\mu\nu}B^{\rho\sigma}.
\end{equation}

From an effective field theory standpoint, there are additional dimension-four operators coupling the KR field to curvature tensors involving the Levi-Civita symbol that we could consider at linear order in the curvature;
\begin{equation}
    \mathcal{L}_{B\tilde{B}R}=\xi_2B^{\mu\lambda}\tilde{B}^\nu_{ \ \ \lambda}R_{\mu\nu}+\xi_3\tilde{B}_{\mu\nu}B^{\mu\nu}R.
\end{equation}
Firstly, these terms are  proportional due to the identity $B_\mu^{ \ \lambda}\tilde{B}_{\lambda\nu}=\frac{1}{4}g_{\mu\nu}B_{\alpha\beta}\tilde{B}^{\alpha\beta},$ but more relevant for our purposes, such operators do not contribute to parity violation. In this work we will therefore focus our attention on (\ref{Lint}).

%=================================================================================================================================
\section{Theory}\label{sec:FullTheory}
%=================================================================================================================================

We consider the full theory given by 
\begin{equation}
    \mathcal{L}=\mathcal{L}_{\text{EH}}+\mathcal{L}_{\text{B}}+\mathcal{L}_{\text{int}}+\mathcal{L}_m,
\end{equation}
where $\mathcal{L}_m$ is any additional matter content and
\begin{equation}
    \mathcal{L}_{\text{EH}}=\frac{m_p^2}{2}R
\end{equation}
is the usual Einstein-Hilbert term, while $\mathcal{L}_{\text{B}}$ and $\mathcal{L}_{\text{int}}$ are given by (\ref{LB}) and (\ref{Lint}), respectively. For an arbitrary background, the field equations for the $B$-field and the metric are given by 
\begin{equation}\label{covEOM}
\begin{split} 
\nabla_\alpha H^{\alpha}_{ \ \ \mu\nu}&= -\frac{\partial V}{\partial B^{\mu\nu}}+4\xi\tilde{R}_{\mu\nu \rho\sigma}B^{\rho\sigma}\\
    m_p^2G_{\mu\nu}&=\xi\Theta_{\mu\nu}+T^{\text{B}}_{\mu\nu}+T_{\mu\nu}^{\text{m}},
    \end{split}
\end{equation}
where $T_{\mu\nu}^{\text{m}}$ is the contribution from the additional matter sector, 
\begin{equation}
    \begin{split} 
    T_{\mu\nu}^{\text{B}}&=-\frac{1}{2}H_\mu^{ \ \ \alpha\beta}H_{\nu\alpha\beta}+g_{\mu\nu}\Big(\frac{1}{12}H_{\alpha\beta\rho}H^{\alpha\beta\rho}-V(B)\Big)
    \end{split}
\end{equation}
is the stress tensor of the free KR field and
\begin{equation} 
\Theta_{\mu\nu}=\nabla^\alpha\nabla^\beta P_{\mu\alpha\nu\beta}-\frac{1}{4}g_{\mu\nu}\tilde{R}_{\alpha\beta\rho\sigma}B^{\alpha\beta}B^{\rho\sigma}
\end{equation}
stems from the new interaction term, where $P$ is given by (\ref{FullP}).

%=================================================================================================================================
\section{GW solution in FLRW background}\label{sec:Parameterization}
%=================================================================================================================================

We will look to make contact with the parameterization given in \cite{Jenks:2023pmk}, where the equations of motion (EOM) for the left/right helicity amplitudes are written
\begin{widetext}
\begin{equation}\label{parameterization}
\begin{split}
   & h''_{\text{R,L}}+\Big\{2\mathcal{H}+\lambda_{\text{R,L}}\sum_{n=1}^{\infty}k^n\Big[\frac{\alpha_n(\eta)}{(\Lambda_{\text{PV}}a)^n}\mathcal{H}+\frac{\beta_n(\eta)}{(\Lambda_{\text{PV}}a)^{n-1}}\Big]\Big\}h'_{\text{R,L}} \\
    &+k^2\Big\{1+\lambda_{\text{R,L}}\sum_{m=0}^{\infty}k^{m-1}\Big[\frac{\gamma_m(\eta)}{(\Lambda_{\text{PV}}a)^m}\mathcal{H}+\frac{\delta_m(\eta)}{(\Lambda_{\text{PV}}a)^{m-1}}\Big]\Big\}h_{\text{R,L}}=0,
\end{split}
\end{equation}
where primes denote partial derivatives with respect to conformal time $\eta$, $\mathcal{H}\equiv a'/a$ is the Hubble parameter, $\lambda_{\text{R,L}}=\pm 1,$ and $\Lambda_{\text{PV}}$ is the EFT cutoff. The summation indices are such that $n$ runs over odd integers and $m$ runs over even integers, and the $\{\alpha_n,\beta_n,\gamma_m,\delta_m\}$ functions generically depend on the conformal time and are theory dependent\footnote{The parameterization written down in \cite{Jenks:2023pmk} focuses solely on parity odd effects. Recently, in \cite{Daniel:2024lev}, the parameterization was extended to include parity even effects, with a focus on Chern-Simons and the Gauss-Bonnet coupling. Similar models were also studied in \cite{Satoh:2007gn,Satoh:2008ck} in the context of string-inspired inflationary scenarios.}.  For metric fluctuations on top of an FLRW cosmology in conformal coordinates, the equations of motion for the gravitational wave amplitudes in our model become
\begin{equation}\label{EOMRL}
\begin{split} 
 \Big(h_{\text{R,L}}''+2\mathcal{H}h_{\text{R,L}}'-\partial^2_zh_{\text{R,L}}\Big)\Big(m_p^2-3\xi\Ve\Vm\Big)+i\lambda_{\text{R,L}}v^2\partial_zh_{\text{R,L}}=0,
    \end{split}
\end{equation}
where $v^2 =\Vm^2-\Ve^2$ is the minimum of the potential. Passing to momentum space by writing $\tilde{h}_{\text{R,L}}(z,\eta)=e^{ikz}\bar{h}_{\text{R,L}}(\eta)$ and simplifying, we obtain
\begin{equation}
h_{\text{R,L}}''+2\mathcal{H}h_{\text{R,L}}'+k^2\Big(1-\lambda_{\text{R,L}}\frac{\mathcal{H}}{k}\frac{\xi \tilde{v}^2}{1-3\xi\tVe\tVm}\Big)h_{\text{R,L}}=0,
\end{equation}
where $\tilde{v}=v/m_p$, $\tVe=\Ve/m_p$, and $\tVm=\Vm/m_p$. From the parameterization (\ref{parameterization}), we can immediately identify 
\begin{equation}
    \begin{split} \gamma_0=-\frac{\xi \tilde{v}^2}{1-3\xi\tVe\tVm}\approx -\xi\tilde{v}^2,
    \end{split} 
\end{equation}
since $\tVe,\tVm\ll 1$. We see that the interaction produces solely birefringence effects in the phase of the gravitational wave polarizations at lowest order. 

\iffalse 
This scenario has the additional interesting feature that the interaction (\ref{Lint}) does not affect the background cosmology. This is apparent from the form of the interaction's contribution to the Einstein equations;
\begin{equation}\label{Thetamunu}
    \Theta_{\mu\nu}=\frac{\Ve\Vm(a''/a-4\mathcal{H}^2)}{a^6}\begin{pmatrix}
        \frac{-4a^2\mathcal{H}^2}{a''/a-4\mathcal{H}^2}&&& \\
        &2&& \\
        &&1& \\
        &&&1
    \end{pmatrix},
\end{equation}
which clearly vanishes when $\Ve=0.$ The cosmological evolution is thus driven by standard matter and the free energy of the KR field, with stress tensor
\begin{equation}
    T_{\mu\nu}^{\text{B}}=-2\Vm^2\mathcal{H}^2\begin{pmatrix}
        1&&& \\
        &1&& \\
        &&-1& \\
        &&&-1
    \end{pmatrix}-a^2V(B)\begin{pmatrix}
        -1&&& \\
        &1&& \\
        &&1& \\
        &&& 1
    \end{pmatrix},
\end{equation}
independent of the value of $\Ve$.
\fi 

Now, the authors of \cite{Jenks:2023pmk} go on to write down the solutions for the polarization modes as a function of redshift $z$ in terms of the parameters of (\ref{parameterization}), which take the general form 
\begin{equation}\label{gensolution}
    h_{\text{R,L}}=h^{\text{GR}}_{\text{R,L}}\text{exp}\Bigg\{\mp\frac{[k(1+z)]^n}{2}\Bigg(\frac{\alpha_{n_0}}{\Lambda_{\text{PV}}^n}z_n+\frac{\beta_{n_0}}{\Lambda_{\text{PV}}^{n-1}}D_{n+1}\Bigg)\Bigg\}\text{exp}\Bigg\{\pm i\frac{[k(1+z)]^m}{2}\Bigg(\frac{\gamma_{m_0}}{\Lambda_{\text{PV}}^m}z_m+\frac{\delta_{m_0}}{\Lambda_{\text{PV}}^{m-1}}D_{m+1}\Bigg)\Bigg\}.
\end{equation}
\end{widetext} 
Here, the summations over $m$ and $n$ are implicit,   $h^{\text{GR}}$ is the standard oscillating solution from GR, $D_\alpha$ and $z_\alpha$ are defined as
\begin{equation}
    \begin{split}
        D_\alpha&=(1+z)^{1-\alpha}\int\frac{(1+z)^{\alpha-2}}{H(z)}dz, \\
        z_\alpha&=(1+z)^{-\alpha}\int\frac{dz}{(1+z)^{1-\alpha}}.
    \end{split}
\end{equation}
The $n_0,m_0$ subscript on $\alpha,\beta,\gamma,\delta$ denote their late time, slowly varying approximations. Our solution has $\gamma_0=-\xi\tilde{v}^2,$ thus only $m=0$ is nontrivial. In the exponent, we therefore have $z_0=\text{ln}(1+z)$, and our solution is then
\begin{equation}\label{sol1}
    h_{\text{R,L}}=h^{\text{GR}}_{\text{R,L}}\text{exp}\Big(\mp \frac{i}{2}\xi\tilde{v}^2z_0\Big).
\end{equation}

As we discussed in Sec. \ref{sec:Intro}, the interaction being a dimension-four operator implies that the coupling $\xi$ is dimensionless, such that there is not a mass scale setting a clear UV cutoff. This manifests in (\ref{sol1}) by way of only an $m=0$ contribution from (\ref{gensolution}) along with $\alpha_n=\delta_m=0$; no factors of $\Lambda_{\text{PV}}$ survive in the parameterization (\ref{parameterization}), although a dimension-four operator in principle could produce $\beta_1\neq 0.$

The dispersion relation for any parity violating theory which can be classified by the parameterization (\ref{parameterization}) is given by equation (30) in \cite{Jenks:2023pmk}. In our case, this becomes
\begin{equation}\label{dispersion}
\begin{split} 
    \omega_{\text{R,L}}^2&=k^2\big\{1+\lambda_{\text{R,L}}k^{-1}\gamma_0\mathcal{H}\big\} \\
    &=k^2\{1+\tilde{\lambda}_{\text{R,L}}k^{-1}\},
\end{split}
\end{equation}
defining 
\begin{equation} \tilde{\lambda}_{\text{R,L}}=-\lambda_{\text{R,L}}(\xi\tilde{v}^2\mathcal{H}).
\end{equation} 
From this we easily obtain the phase $v_p=\omega/k$ and group $v_d=d\omega/dk$ velocities,
\begin{equation}\label{velocities1}
    \begin{split} 
    v_p&=\sqrt{1+\tilde{\lambda}_{\text{R,L}}k^{-1}}, \\
    v_g&=\frac{1+\tfrac{1}{2}\tilde{\lambda}_{\text{R,L}}k^{-1}}{\sqrt{1+\tilde{\lambda}_{\text{R,L}}k^{-1}}},
    \end{split}
\end{equation}
which are clearly different between $h_{\text{R}}$ and $h_{\text{L}}$. Taking a closer look at the group velocity, we have
\begin{equation}\label{vg}
    \begin{split}
        v_g&=(1+\tfrac{1}{2}\tilde{\lambda}_{\text{R,L}}k^{-1})(1-\tfrac{1}{2}\tilde{\lambda}_{\text{R,L}}k^{-1}+\tfrac{3}{8}\tilde{\lambda}_{\text{R,L}}^2k^{-2}+\cdots) \\
        &\simeq 1+\tfrac{1}{8}\tilde{\lambda}_{\text{R,L}}^2k^{-2}+O\Big(\tilde{\lambda}_{\text{R,L}}^3k^{-3}\Big)
    \end{split}
\end{equation}
for $(\xi\tilde{v}^2\mathcal{H})\ll 1.$ Since the correction appears at $O(\tilde{\lambda}^2)$, the sign difference between the left and right handed amplitudes squares away such that they propagate with equal group velocity to leading order. And because the correction is non-negative, the model predicts a propagation speed slightly larger than one. Comparing to the phase velocity, we see that 
\begin{equation}
    v_p\simeq 1+\tfrac{1}{2}\tilde{\lambda}_{\text{R,L}}k^{-1}-\tfrac{1}{8}\tilde{\lambda}_{\text{R,L}}^2k^{-2}+O\Big(\tilde{\lambda}_{\text{R,L}}^3k^{-3}\Big).
\end{equation}
The correction to the phase velocity due to the parity violating interaction clearly appears at a lower order than the group velocity, and the sign difference between $h_{\text{R}}$ and $h_{\text{L}}$ survives.

%=================================================================================================================================
\section{Comments on constraints}\label{sec:Constraints}
%=================================================================================================================================
The possibility of parity violation to be observed in gravitational wave detectors was suggested well over 10 years ago \cite{Alexander:2007kv,Yunes:2010yf}. The data currently being gathered from merger processes \cite{LIGOScientific:2014pky,VIRGO:2014yos,LIGOScientific:2020ibl,KAGRA:2021vkt} can in principle constrain parity violating parameters, although much more progress on this front will be made from third generation detectors due to their increased sensitivities \cite{Maggiore:2019uih,Evans:2021gyd}. Additionally, very recently, an analysis to constrain parity violation was carried out in \cite{Callister:2023tws} using data from pulsar timing arrays \cite{NANOGrav:2023gor,NANOGrav:2023hvm,EPTA:2023fyk,Xu:2023wog,Reardon:2023gzh}. The two primary effects of interest are velocity birefringence and amplitude birefringence. The former refers to right- and left-handed waves propagating with different frequency dependent phase shifts, the latter corresponding to an attenuation or amplification of one polarization over another. It is also the case that modified theories of gravity generally predict a propagation speed different from unity. In what follows, we will focus on propagation speed.

%=================================================================================================================================
\subsection{Propagation speed}
%=================================================================================================================================

The observation of the coincident gamma ray burst (GRB)/gravitational wave signal GW170817 has constrained the GW propagation speed to be \cite{LIGOScientific:2017zic} 
\begin{equation}\label{cgwConstraint}
    -3\times 10^{-15}<c_{\text{gw}}-1<7\times 10^{-16}.
\end{equation}
The measurement was used to constrain parity violating effects in ghost-free scalar tensor models in \cite{Nishizawa:2018srh} and in symmetric teleparallel gravity in \cite{Conroy:2019ibo}. Using our result for the group velocity (\ref{vg}), this translates to 
a bound on the parameter $\tilde{v}=v/m_p$. If we take the dimensionless coupling to be $\xi\sim O(1)$, and use the upper bound on (\ref{cgwConstraint}), then
\begin{equation}
    \frac{v^4}{m_p^4}\Big(\frac{\mathcal{H}}{k}\Big)^2=\frac{v^4}{m_p^4}\Big(\frac{H_0}{ka}\Big)^2<5.6\times 10^{-15},
\end{equation}
where $H_0$ is the Hubble constant and $a=a(t)$ is the scale factor in cosmic time. Using the SH0ES result \cite{Riess:2021jrx} for the Hubble constant $H_0\approx 73.04$ km/s/Mpc, we can write this as a function of redshift $z$ and the gravitational wave frequency $f$ as
\begin{equation}\label{vconstraint}
\frac{v}{m_p}\lesssim 4.4\times 10^5\Bigg(\frac{f/\text{Hz}}{1+z}\Bigg)^{1/2}.
\end{equation}
This illustrates that observations of lower frequency gravitational waves at high redshift would give the most meaningful constraint on $v$. This is particularly challenging, however, since the lowest frequency gravitational waves observed in pulsar timing arrays are from the stochastic GW background \cite{NANOGrav:2023gor}. Since the source of the background is not yet known, redshift determination is not feasible. If the GW background is determined to be dominantly caused by the physics of inflation, (\ref{vconstraint}) will be useful in constraining the energy scale of the KR fields\footnote{We note that a very similar model including couplings to the Riemann tensor (opposed to its dual) and the Ricci tensor will indeed give an explicit constraint of $v\lesssim 10^{11}$ GeV \cite{ToAppear}. }.

\section{Discussion}\label{sec:Discussion} Third generation gravitational wave detectors such as the Einstein Telescope \cite{Maggiore:2019uih} and Cosmic Explorer \cite{Evans:2021gyd} are anticipated to have sensitivities capable of probing beyond GR effects such as parity violation \cite{Califano:2023aji}. Parity violation may also be observed in large scale structure \cite{Vanzan:2023cze,Creque-Sarbinowski:2023wmb}. Any observation definitively signaling parity violation would be a smoking gun for gravitational physics beyond General Relativity. It is therefore of great interest to explore various models which predict gravitational parity violation in anticipation of such data. In this work, inspired by recent attention on the Kalb-Ramond field as a dark matter candidate, we considered parity violation from a dimension-four operator  coupling the dual Riemann tensor to the KR fields, $\mathcal{L}_{\text{int}}=-\tfrac{\xi}{2}\tilde{R}_{\mu\nu\rho\sigma}B^{\mu\nu}B^{\rho\sigma}$. For simplicity, we constructed a scenario where the KR fields can be treated as being constant. This will happen provided the KR field experiences a potential that satisfies $\frac{\partial V}{\partial B^{\mu\nu}}\big|_{B^{(0)}_{\mu\nu}}=0$ for $B^{(0)}_{\mu\nu}\neq 0.$ Two parameters $\Ve,\Vm$, characterize the degrees of freedom of $B_{\mu\nu}.$ After writing down the field equations for the gravitational waves in the circular basis for arbitrary $\Vm,\Ve,$ we mapped the solution to the general parameterization presented in \cite{Jenks:2023pmk}, which depends solely on $v=\sqrt{\Vm^2-\Ve^2}$ at lowest order in $\Vm/m_p,\Ve/m_p$. Constraints from the GRB/gravitational wave signal GW170817 \cite{LIGOScientific:2017zic} produced the bound given by (\ref{vconstraint}). %As we discussed in Sec. \ref{sec:AmpBi}, this upper bound for dark matter being near the PeV scale is interesting from  theoretical \cite{Wells:2004di,Bhattiprolu:2023lfh} and phenomenological \cite{TibetASgamma:2021tpz,Maity:2021umk} standpoints.

We assumed a generic potential was responsible for the KR field settling into its background value, which triggers spontaneous Lorentz violation. As a result, (pseudo-) Nambu-Goldstone modes associated to the field fluctuations along the broken generators emerge \cite{Bluhm:2004ep,Bluhm:2007bd,Higashijima:2001sq,Hernaski:2016dyk}. Consequences of this spontaneous symmetry breaking were studied extensively in \cite{Altschul:2009ae}, where the authors derived reasonable constraints on the Lorentz violating coefficients  present in certain non-minimal curvature couplings (see, e.g. \cite{Cheng:2006us,Carroll:2004ai,Lambiase:2005kb,Colladay:1998fq,Kostelecky:1988zi,Kostelecky:2008ts,Kostelecky:2003fs,Amarilo:2023wpn}). However, such modes do not play an important role in our work here. 

\iffalse 
To get a feel for where the detectability of this model lies, we can consider the polarization degree 
$$P(k)=\frac{\langle h_{\text{R}}(k)h^*_{\text{R}}(k') -h_{\text{L}}(k)h_{\text{L}}^*(k')\rangle}{\langle h_{\text{R}}(k)h_{\text{R}}^*(k')+h_{\text{L}}(k)h_{\text{L}}^*(k')\rangle},$$
where the brackets denote a spatial averaging \cite{Kahniashvili:2020jgm,Ellis:2020uid,Li:2024fxy}. In our simplified scenario of a single
wave propagating in the $z$-direction, we have $P(k)=-\tanh(\xi\tilde{\mathcal{V}}_m^2kD_C/2)$. Expanding the dispersion relation (\ref{dispersion}) and working to lowest order in $\tV_m$, we can write the polarization degree as function of frequency, $P(\omega)\approx\frac{1}{2}\xi\tV_m^2\omega D_C$. Saturating $\tV_m$ at its upper bound (\ref{Vmax}) and taking $\xi\sim 1$, we find $P(\omega)\sim10^{-22}\omega D_C$. Detecting a nonzero polarization degree clearly favors a higher frequency signal, corresponding to lighter masses involved in sourcing the gravitational wave emission. Relatively low mass merger events have been observed \cite{LIGOScientific:2024elc} and are expected to have interesting multimessenger phenomenology \cite{Matur:2024nwi}.
\fi

Since we only observe velocity birenfrigence, detecting the parity violation from this simple model relies heavily on multimessenger signals a l\'a (\ref{vconstraint}), whereas amplitude birefringence can be detected without the need for such a signal. In either case, definitively detecting gravitational wave birefringence is a formidable task that naturally depends on the sensitivities of the detectors \cite{Moore:2014lga}. Significant efforts have gone into studying detection prospects in the stochastic GW background \cite{Seto:2006hf,Kato:2015bye,Smith:2016jqs,Callister:2017ocg,Domcke:2019zls,Omiya:2020fvw,LIGOScientific:2020tif,Wu:2021kmd,Sato-Polito:2021efu,Omiya:2023rhj,Mentasti:2023gmg,AnilKumar:2023hza,Callister:2023tws}, in addition to analysis focusing on late-time merger events. In the event that a conclusive observation of birefringence is made in, say, detections from merger processes, we will not explicitly know which scenario was responsible provided the signal's properties can satisfactorily be described within known bounds on the model parameters. We expect that discerning the correct model will only be possible if its (beyond standard cosmology) predictions are verified at different scales and in different physical systems. For example, one would need precision measurements from next generation CMB observatories, and perhaps primordial gravitational wave precision properties, to discriminate between models that in some systems, predict the same physics.

There are numerous next directions to pursue. On one hand, couplings to the Riemann tensor contribute nontrivially to gravitational entropy \cite{Parikh:2009qs,LopesCardoso:1999cv}. Many parity violating theories, such as the interaction we studied in this work (\ref{Lint2}) and better studied theories such as dynamical Chern-Simons gravity, entail couplings to the Riemann tensor and therefore will modify the Wald entropy \cite{Wald:1993nt}. Entanglement entropy in quantum field theory was studied in the context of parity violating theories in \cite{Hughes:2015ora}, where the authors made contact with gravitational anomalies in flat space. It would be interesting to explore a connection between gravitational entropy corrections and parity violation in black hole physics. More generally, black hole solutions with KR hair modify the shadow and lensing properties in comparison to GR \cite{Kumar:2020hgm,Lessa:2019bgi}. Repeating such an analysis would offer a different approach to constraining our model parameters, as we certainly expect additional modifications as a result of the interaction (\ref{Lint}).

It is of significant importance to consider the consequences of the interaction (\ref{Lint}) in early universe physics. The KR field and more general $p$-forms in inflationary scenarios have been been studied in Starobinsky inflation \cite{Elizalde:2018now}, $F(R)$ theories \cite{Elizalde:2018rmz}, anisotropic inflation \cite{Almeida:2019xzt}, and in general cosmological backgrounds \cite{BeltranAlmeida:2018nin}. Each of these works assumed even parity and gauge invariance. Gauge invariance being preserved implies the KR field can be described in terms of one (pseudo)scalar degree of freedom, which is not an approach that we can take due to the non gauge invariance of (\ref{Lint}). It is therefore interesting to explore the consequences of (\ref{Lint}) in inflationary scenarios, in particular the deviations from GR for the primordial gravitational wave spectrum as explored in \cite{Kahniashvili:2020jgm,Li:2024fxy,Ellis:2020uid}, in addition to consistency with other cosmological observables (for example, parity violation in the CMB \cite{Gaztaaga2024:2401.08288v1}). Allowing the KR fields to be fully dynamical opposed to frozen in a constant background value will be relevant in such scenarios. One approach would be to study the interactions with gravity in their polarization basis, as written in \cite{Capanelli:2023uwv}. On the other hand, as mentioned in Section \ref{sec:KRfeatures}, genuinely preserving cosmological isotropy when the KR field is in its background value should necessitate multiple fields, which may result in modifications to our results in this paper.

%The simplified scenario of $\mathcal{V}_e=0$ resulted in only parity-odd contributions to the GW propagation, and additionally implied the interaction term (\ref{Lint}) did not affect the background evolution of the spacetime, as seen from (\ref{Thetamunu}). The more general case $\mathcal{V}_e\neq 0$ contributes parity even and parity odd corrections to the GW propagation, and therefore can be mapped to the parameterization recently presented in \cite{Daniel:2024lev}.

It will be interesting to see whether or not considering constraints from parity-even beyond GR models can tighten the upper bound on energy scale for the KR field, $v=\sqrt{\mathcal{V}_m^2-\mathcal{V}_e^2}$. This scenario also nontrivially affects the background evolution, which, in the context of early universe physics, can have a drastic affect on cosmological observables. A complete analysis will further shed light on the viability of non-minimal dimension-four couplings between the Riemann tensor and the KR field. We leave this and other explorations to future work.

\begin{center}
{\bf ACKNOWLEDGEMENTS }
\end{center}
 The authors thank Heliudson Bernardo, Tatsuya Daniel, Leah Jenks, Evan McDonough, and Andrew Svesko for helpful comments and discussions. SA and TM are supported by the Simons Foundation, Award 896696.

%\clearpage

%\nocite*

\bibliographystyle{hunsrt2}
\bibliography{bib}
\end{document}